\documentstyle[twoside]{article}

\catcode`\@=11
\long\def\@makefntext#1{ 
\protect\noindent \hbox to 3.2pt {\hskip-.9pt
$^{{\eightrm\@thefnmark}}$\hfil}#1\hfill} 

\def\thefootnote{\fnsymbol{footnote}}
 \def\@makefnmark{\hbox to 0pt{$^{\@thefnmark}$\hss}}  

\def\ps@myheadings{\let\@mkboth\@gobbletwo
\def\@oddhead{\hbox{} 
\rightmark\hfil\eightrm\thepage}
\def\@oddfoot{}\def\@evenhead{\eightrm\thepage\hfil 
\leftmark\hbox{}}\def\@evenfoot{}
\def\sectionmark##1{}\def\subsectionmark##1{}}

\oddsidemargin=\evensidemargin
\renewcommand{\thefootnote}{\fnsymbol{footnote}}

\newcounter{sectionc}\newcounter{subsectionc}\newcounter{subsubsectionc}
\renewcommand{\section}[1] {\vspace{12pt}\addtocounter{sectionc}{1}
\setcounter{subsectionc}{0}\setcounter{subsubsectionc}{0}\noindent
        {\tenbf\thesectionc. #1}\par\vspace{5pt}}
\renewcommand{\subsection}[1] {\vspace{12pt}\addtocounter{subsectionc}{1}
        \setcounter{subsubsectionc}{0}\noindent
        {\bf\thesectionc.\thesubsectionc. {\kern1pt \bfit #1}}\par\vspace{5pt}}
\renewcommand{\subsubsection}[1]
{\vspace{12pt}\addtocounter{subsubsectionc}{1}
        \noindent{\tenrm\thesectionc.\thesubsectionc.\thesubsubsectionc.
        {\kern1pt \tenit #1}}\par\vspace{5pt}}
\newcommand{\nonumsection}[1] {\vspace{12pt}\noindent{\tenbf #1}
        \par\vspace{5pt}}

\newcounter{appendixc}
\newcounter{subappendixc}[appendixc]
\newcounter{subsubappendixc}[subappendixc]
\renewcommand{\thesubappendixc}{\Alph{appendixc}.\arabic{subappendixc}}
\renewcommand{\thesubsubappendixc}
        {\Alph{appendixc}.\arabic{subappendixc}.\arabic{subsubappendixc}}

\renewcommand{\appendix}[1] {\vspace{12pt}
        \refstepcounter{appendixc}
        \setcounter{figure}{0}
        \setcounter{table}{0}
        \setcounter{lemma}{0}
        \setcounter{theorem}{0}
        \setcounter{corollary}{0}
        \setcounter{definition}{0}
        \setcounter{equation}{0}
        \renewcommand{\thefigure}{\Alph{appendixc}.\arabic{figure}}
        \renewcommand{\thetable}{\Alph{appendixc}.\arabic{table}}
        \renewcommand{\theappendixc}{\Alph{appendixc}}
        \renewcommand{\thelemma}{\Alph{appendixc}.\arabic{lemma}}
        \renewcommand{\thetheorem}{\Alph{appendixc}.\arabic{theorem}}
        \renewcommand{\thedefinition}{\Alph{appendixc}.\arabic{definition}}
        \renewcommand{\thecorollary}{\Alph{appendixc}.\arabic{corollary}}
        \renewcommand{\theequation}{\Alph{appendixc}.\arabic{equation}}
        \noindent{\tenbf Appendix \theappendixc #1}\par\vspace{5pt}}
\newcommand{\subappendix}[1] {\vspace{12pt}
        \refstepcounter{subappendixc}
        \noindent{\bf Appendix \thesubappendixc. {\kern1pt \bfit #1}}
        \par\vspace{5pt}}
\newcommand{\subsubappendix}[1] {\vspace{12pt}
        \refstepcounter{subsubappendixc}
        \noindent{\rm Appendix \thesubsubappendixc. {\kern1pt \tenit #1}}
        \par\vspace{5pt}}

\newcommand{\textlineskip}{\baselineskip=13pt}
\newcommand{\smalllineskip}{\baselineskip=10pt}

\def\eightcirc{
\begin{picture}(0,0)
\put(4.4,1.8){\circle{6.5}}
\end{picture}}
\def\eightcopyright{\eightcirc\kern2.7pt\hbox{\eightrm c}}

\newcommand{\copyrightheading}[1]
        {\vspace*{-2.5cm}\smalllineskip{\flushleft
        {\eightrm Modern Physics Letters A, #1}\\
        {\eightrm $\eightcopyright$\, World Scientific Publishing
         Company}\\
         }}

\newcommand{\publisher}[2]{{\begin{center}\eightrm\smalllineskip
        Received #1\\
        Revised #2
        \end{center}
        }}

\def\abstracts#1#2#3{{
        \centering{\begin{minipage}{4.5in}\baselineskip=10pt\eightrm
        \centerline{ABSTRACT}
        \parindent=0pt #1\par
        \parindent=15pt #2\par
        \parindent=15pt #3
        \end{minipage} }\par}}

\renewenvironment{thebibliography}[1]                   
        {\ninerm
         \baselineskip=11pt                             
         \begin{list}{\arabic{enumi}.}
        {\usecounter{enumi}\setlength{\parsep}{0pt}
         \setlength{\leftmargin 17pt}{\rightmargin 0pt} 
         \setlength{\itemsep}{0pt} \settowidth          
        {\labelwidth}{#1.}\sloppy}}{\end{list}}

\newcounter{itemlistc}
\newcounter{romanlistc}
\newcounter{alphlistc}
\newcounter{arabiclistc}

\newcommand{\fcaption}[1]{
        \refstepcounter{figure}
        \setbox\@tempboxa = \hbox{\eightrm Fig.~\thefigure. #1}
        \ifdim \wd\@tempboxa > 5in
           {\begin{center}
        \parbox{5in}{\eightrm \smalllineskip Fig.~\thefigure. #1 }
            \end{center}}
        \else
             {\begin{center}
             {\eightrm Fig.~\thefigure. #1}
              \end{center}}
        \fi}

\newcommand{\tcaption}[1]{
        \refstepcounter{table}
        \setbox\@tempboxa = \hbox{\eightrm Table~\thetable. #1}
        \ifdim \wd\@tempboxa > 5in
           {\begin{center}
        \parbox{5in}{\eightrm\smalllineskip Table~\thetable. #1 }
            \end{center}}
        \else
             {\begin{center}
             {\eightrm Table~\thetable. #1}
              \end{center}}
        \fi}

\def\@citex[#1]#2{\if@filesw\immediate\write\@auxout    
        {\string\citation{#2}}\fi                       
\def\@citea{}\@cite{\@for\@citeb:=#2\do                 
        {\@citea\def\@citea{,}\@ifundefined             
        {b@\@citeb}{{\bf ?}\@warning
        {Citation `\@citeb' on page \thepage \space undefined}}
        {\csname b@\@citeb\endcsname}}}{#1}}

\newif\if@cghi
\def\cite{\@cghitrue\@ifnextchar [{\@tempswatrue
        \@citex}{\@tempswafalse\@citex[]}}
\def\citelow{\@cghifalse\@ifnextchar [{\@tempswatrue
        \@citex}{\@tempswafalse\@citex[]}}
\def\@cite#1#2{{$\null^{#1}$\if@tempswa\typeout
        {IJCGA warning: optional citation argument
        ignored: `#2'} \fi}}

\def\pmb#1{\setbox0=\hbox{#1}
        \kern-.025em\copy0\kern-\wd0
        \kern.05em\copy0\kern-\wd0
        \kern-.025em\raise.0433em\box0}

\def\fnt#1#2{\footnotetext{\kern-.3em
        {$^{\mbox{\scriptsize #1}}$}{#2}}}

\def\fpage#1{\begingroup
\voffset=.3in
\thispagestyle{empty}\begin{table}[b]\centerline{\footnotesize #1}
        \end{table}\endgroup}

\def\runninghead#1#2{\pagestyle{myheadings}
\markboth{{\eightit{\quad #1}}\hfill}{\hfill{\eightit{#2\quad}}}}

\font\tenbf=cmbx10
\font\tenit=cmti10
\font\tenit=cmti10
\font\bfit=cmbxti10 at 10pt
 1
 1
 1

\font\ninerm=cmr9

\font\eightrm=cmr8
\font\eightit=cmti8

\def\qed{\hbox{${\vcenter{\vbox{                          
   \hrule height 0.4pt\hbox{\vrule width 0.4pt height 6pt
   \kern5pt\vrule width 0.4pt}\hrule height 0.4pt}}}$}}

\runninghead{Mathematical Remarks on the  Cohomology of
$\ldots$} {Mathematical Remarks on the  Cohomology of
$\ldots$}

\def\A{{\cal A}}
\def\G{{\cal G}}
\def\R{{\bf R}}
\def\C{{\bf C}}
\def\Map{{\rm Map}}
\def\Z{{\bf Z}}

\begin{document}
\normalsize\textlineskip
{\thispagestyle{empty}
\setcounter{page}{1}

\renewcommand{\thefootnote}{\fnsymbol{footnote}} 

\copyrightheading{Vol. 0, No. 0 (1992) 000--000}

\vspace*{0.88truein}

\fpage{1}
\centerline{\bf MATHEMATICAL REMARKS ON THE }
\vspace*{0.035truein}
\centerline{\bf COHOMOLOGY OF GAUGE GROUPS AND ANOMALIES}
\vspace{0.37truein}
\centerline{\footnotesize A.L. CAREY}
\vspace*{0.015truein}
\centerline{\footnotesize\it Department of Pure Mathematics, University of
Adelaide}
\baselineskip=10pt
\centerline{\footnotesize\it Adelaide SA 5005, AUSTRALIA.}
\vglue 10pt
\centerline{\eightrm M.K. MURRAY}
\vspace*{0.015truein}
\centerline{\footnotesize\it Department of Pure Mathematics, University of
Adelaide}
\baselineskip=10pt
\centerline{\footnotesize\it Adelaide SA 5005, AUSTRALIA.}
\vspace{0.225truein}
\publisher{(received date)}{(revised date)}

\vspace*{0.21truein}
\abstracts{\noindent
 }{Anomalies can be viewed as arising from the cohomology of the Lie
algebra of the group of gauge transformations and also from the topological
cohomology of the group of connections modulo gauge transformations.
We show how these two approaches are unified by the transgression map.
We discuss the geometry behind the current commutator anomaly and the Faddeev-
Mickelsson anomaly using the recent notion of a gerbe. Some anomalies
(notably 3-cocycles) do not have such a geometric
origin. We discuss one example and a conjecture on how these may be related to
geometric anomalies}{}

}
\textheight=7.8truein
\setcounter{footnote}{0}
\renewcommand{\thefootnote}{\alph{footnote}}

\section{Introduction and preliminaries}
\noindent
The transition from classical to quantum physics has lead to a great deal
of interesting
mathematics. The study of so-called anomalies
really begins with Wigner's theorem: symmetries of a dynamical system
manifest themselves as projective representations at the quantum level
and hence give rise to 2-cocycles on the symmetry group.
In quantum field theory unexpected or anomalous behaviour of gauge symmetry
groups
(infinite dimensional Lie groups) in the quantised theory has led to an
extensive study of Lie group cohomology and to the explicit construction
of cocycles (the so-called anomalies). However the infinite dimensional
theory exhibits a number of
features which have no counterpart in finite dimensions. Necessarily new tools
are needed to study the mathematical questions raised by these developments.
In this note we will put forward a particular viewpoint about
`anomalies'.
There is nothing in this paper which is particularly deep.
Our aim is to expose some mathematical questions,
answer some of the easier ones and review or explain related work of
other authors in our framework.

In the first section we define the transgression map. This relates the
Lie algebra cohomology of $L\G$, the Lie algebra of the
gauge group $\G$,  with values in real-valued functions on $\A$,
in dimension $p$ to the
topological cohomology of $\A/\G$, the space of connections modulo
gauge,  in dimension $p+1$.
We then consider particular cases of the transgression map and discuss the
geometric  interpretation of this cohomology and
the corresponding anomaly in physics.  When $p=1$,
$H^2(\A/\G, \Z)$ classifies the equivalence classes of complex,
one-dimensional vector  bundles on $\A/\G$. The anomaly of interest is the
current
commutator anomaly
and is associated to the determinant line bundle of the
Dirac operator.
When $p=2$ the anomaly of interest is the  Faddeev-Mickelsson anomaly.
Only recently
has it become clear the geometric object in this case is what is known as a
gerbe. As gerbes are rather recent phenomena we
present a short review of their properties.
Finally when $p=3$ we give an example of a $3$-cocycle which arises as
an obstruction to the existence of an extension of one Lie algebra by
another.
It is an open question whether it has a relationship with $H^4(\A/\G, \Z)$.

\subsection{Some mathematical preliminaries}
\noindent
Let $M$ be a compact Riemannian manifold and let $P$
be a principal bundle over $M$ with structure group $G$, a compact Lie group.
We let $\A$ denote the affine space of connections on $P$ with values
in the Lie algebra  $LG$ of $G$. Denote by $\G$ the
group of automorphisms of $P$ fixing the base.
By suitable basepointing we can assume that
$\G$ acts freely on $\A$.\cite{singer}
Denote by $L\G$ the Lie algebra of $\G$.

We denote the space of  smooth $p$ forms on a manifold  $X$ by
$\Omega^p(X)$  and  de Rham complex by $\Omega^*(X)$.
If $ H$ is an $LG$-module then we denote  Lie algebra cochains on
$LG$ of degree $p$
with values in $H$ by $C^{p}({LG}, {H})$.
We denote the coboundary operator for this cochain complex by $\delta$ and
the cohomology by $H^*_{LG}(H)$.

If $X$ is a space on which $G$ acts freely then there is a map from
the de Rham cohomology of $X$ into the Lie algebra cohomology of
$\Map(X, R)$ defined as follows. If $\omega$ is a differential $p$
form we can define a Lie algebra cocycle $c_\omega$ by taking $p$ elements
$\eta_i$
of $LG$ and considering the vector fields $(\eta_i)_X$ they induce on
$X$. Then define $c_\omega$ by
$$
c_\omega(\eta_1, \dots, \eta_p)(x) = \omega_x((\eta_1)_X, \dots , (\eta_p)_X).
$$
It is straightforward to check that $c_{d\omega} = \delta(c_\omega)$ where
$\delta$ is the Lie algbra coboundary map. It follows that this defines a map
on cohomology.

\section{The Transgression Map}
\noindent
One can think of anomalies as arising from  the topology
of $\A/\G$ or  as group cocycles on $L \G$  and both these points
of view have been explored in the literature to some extent. We shall show
that they
are related via the {\em transgression} map.

If $\omega\in\Omega^{p}(\A/\G)$, $d\omega=0$ and $\pi:\A
\rightarrow\A/\G$ then $\pi^*\omega=d\mu$ for some $\mu$ as
$\A$ is contractible.  In fact there will be an explicit formula for
$\mu$ as $\A$ has an affine structure.   Consider $c_\mu$  the $LG$ cocycle
obtained by restricting $\mu$ to the tangent spaces to the fibres of $\A
\to \A/\G$
as in section 1.1. Then $\delta(c_\mu) = c_{d\mu} = c_\omega$.  But
$\omega$ is pulled-back so  kills any vectors tangent to the  fibres and
hence $c_\omega = 0$.   Hence $\delta(c_\mu) = 0$.
Moreover if we change $\mu$ by $\mu+d\eta$ (the only possible change because
$\A$ is contractible) then $\mu$ changes by $\delta\mu$.
Lastly if we add to $\omega$ some $d\rho$ then this adds $\pi^*\rho$ to
$\mu$ and $\pi^*\rho$ vanishes in the fibre direction {\em unless $\rho$
is a function}.

So have we the {\em transgression map} on  de Rham cohomology:
$$
H^p(\A/\G) \to H^{p-1}_{L\G}(\Map(\A, \R)) \quad\quad p>1
\eqno (1.1)
$$
Notice this is slightly different to the topologists transgression map, they
look at only one fibre and identify it with $\G$ to obtain
$$
H^p(\A/\G) \to  H^{p-1}(\G).
$$

The transgression map relates the cohomology of $\A/\G$ to the Lie algebra
cohomology of the gauge group. In the next two sections we wish to examine
two cases of the transgression map which give rise to anomalies in
physics and which also have interesting geometric interpretations.

\section{ The  transgression for $p=1$ and the determinant line bundle.}
\noindent
Consider a $U(1)$ bundle $Q \to \A$ on which $\G$
acts covering the action of $\G$ on $\A$.  This means that
we can form the quotient $Q/\G$ which is a $U(1)$ bundle
on $\A/\G$.    As a bundle on $\A$,
$Q$ is trivial because $\A$ is contractible. The question we wish
to consider is whether or not $Q/\G$ is trivial. There are two
ways of aproaching this.  We can ask whether $Q$ has a $\G$ equivariant
section. If it has this will descend to a section of $Q/\G$ and hence
trivialise it. Conversely any section of $Q/\G$ will lift back to a
$\G$ equivariant section of $Q$.  The second point of view is
to consider the Chern class of $Q/\G$, this is a cohomology class
in $H^2(\A/\G, \R)$ and vanishes precisely when $Q/\G$ is trivial.
We will show that these two different points of view are related
by the transgression map.

Let $s$ be a section of $Q$ over $\A$ and define $M(A,g)$ by
$$
s(A,g)g^{-1}=M(A,g)s(A).
$$
We consider $M$ as a $\G$ cocycle with values in
$\Map(\A, U(1))$. It measures the failure of $s$ to be
a $\G$ equivariant section. Now
$$
s(Agh)(gh)^{-1}=M(A,gh)s(A)
$$
and
$$
s((Ag)h)h^{-1}g^{-1} = M(Ag,h)s(Ag)g^{-1}
 = M(Ag,h)M(A,g)s(A)
$$
so that
$$
M(A,gh)=M(Ag,h)M(A,g)
$$
or
$$
M(Ag,h)M(A,gh)^{-1}M(A,g)=1
$$
and $M$ is a 1-cocycle for $\G$ with values in $\Map(\A, U(1))$.

Of course even if $Q$ admits a $\G$ equivariant section $s$ may not
be it.  However any other section is of the form $s(A)h(A)$ for
$h \in \Map(\A, U(1))$ and it gives rise to a $M'$ which is related
to $M$ by
$$
M(A, g) = h(Ag)M'(A,g)h(A).
$$
So we see from this that the class of $M$ depends only on the bundle $Q$ and
moreover that if $Q$ admits a $G$ equivariant
section there is an  $h$ such that
$M(A, g) = h(Ag)h(A)$. That is, $M$ is the trivial class.
Conversely if $M(Ag)=h(Ag)h(A)^{-1}$ for some $h:\A\to U(1)$,
then $\widetilde{s}=s(A)h(A)^{-1}$ satisfies
$$
\begin{array}{lcl}
\widetilde{s}(Ag)g^{-1} & = & s(Ag)h(Ag)^{-1}g^{-1} \\[+5pt]
 & = & M(A,g)s(A)M(A,g)^{-1}h(A)^{-1} \\[+5pt]
 & = & \widetilde{s}(A).
\end{array}
$$
So the cohomology class of $M$ is precisely the obstruction to $ Q$ having
a $\G$
invariant trivialization.

To relate this to transgression we consider the Lie algebra version.
Let $\eta$ be an element of  $LG$, then it gives rise to a
vector field $\eta_\A$ on $\A$. There are two ways of lifting this
vector field to the bundle $Q$.  The first way is to note
that by assumption $\G$ acts on $Q$ and hence $\eta$
gives rise to a vector field $\eta_Q$. The second is $s_*(\eta_\A)$,
the lift of  $\eta_\A$ using the section $s$. If $s$ is
equivariant these coincide so we define
$$
m(A,\eta)=s_{*}(\eta)-\hat{\eta}.
$$
Recall that the vertical vectors to the fibering $Q \to \A$
can be identified with the Lie algebra of $U(1)$  and that
we identify with $\R$ so we will regard, $m$
via these identifications as $\R$ valued.
Note that this is related to the group cocycle $M$ by
$$
m(A,\eta)=\left.{\frac{d}{dt}\;\big(M(A,e^{t\eta})\big)}\right|_{t=0}.
$$

It is straightforward to show that $m$ is a Lie algebra
cocycle with values in $\Map(\A, \R)$.  We want to show that $m$
is the transgression of the chern class of $Q$. To see this
pick a connection $\omega$ for $Q/\G$, this
is a $1$-form on $Q/\G$, and let $\pi^*(\nabla)$
be its lift to $Q$.  Denote by $F$ the curvature of $\nabla$, a two-form
on $\A/\G$. The curvature of $\pi^*(\nabla)$ is $\pi^*(F)$.
We will show that the transgression of $F$ is $m$. To
see this notice that on the total space of the bundle $Q$
we have
$$
p^*\pi^*F = d\pi^*(\omega)
$$
where $p \colon Q \to A$ is the
bundle projection. If we pull back with $s$ we have
$$
\pi^*(F) = d s^*\pi^*(\omega)
$$
so that we can take $\mu = s^*\pi^*(\omega)$ in the
definition of  the transgression of $F$.  Let $\eta_\A$ then be the vector
field generated by $\eta$ on $\A$ and similarly for $\eta_Q$.
Notice that $\pi^*(\omega)(\eta_Q) = 0$ as  $\eta_Q$ is
vertical for the projection $Q \to Q/\G$ and $\pi^*(\omega)$
is pulled-back. Also $s_*(\eta) - \eta_Q$ is vertical for the fibering
of $Q\to \A$ so that we have
$$
\pi^*(\omega)(s_*(\eta) - \eta_Q) = s_*(\eta) - \eta_Q.
$$
Combining these facts gives
$$
\mu(\eta_\A) = \pi^*(\omega)(s_*(\eta) = \pi^*(\omega)(s_*(\eta) - \eta_Q) =
s_*(\eta) - \eta_Q = m(A, \eta)
$$
as required.
Notice that if {\em $\G$ is connected} $m$ is precisely the
obstruction to $Q$ having a $\G$ invariant trivialization.

\subsection{The determinant line bundle and the current commutator anomaly.}
\noindent
The anomaly that arises in this case is called the current
commutator anomaly and is related
to the determinant line bundle.  Recall that if $X: V \to W$
is linear map and $\dim(W) = \dim(V)$
then $\det(X)$ is not a number but an element of
$\det(V)^* \otimes \det(W)$ where $\det(V) $ is the highest
exterior power of $V$ and similarly for $W$. Such a situation
arises for the Dirac operator when $M$ is odd dimensional.
Then the Dirac operator maps from plus to minus spinors.
If we couple the Dirac operator to a connection in $\A$ then
we have a family of operators
$
D_A \colon H^+ \to H^-
$
where $H^\pm$ is the space of $\pm$ spinor fields. To
define the determinant of the Dirac operator in this case we have the added
problem of `regularising' the infinite determinant. This can be done
but the regularisation depends on $A$.\cite{Freed} The upshot
is that $\det(D_A)$ is a  section of a line bundle $Q$ over $\A$.
If we want $\det(D_A)$ to be a number we need to be able to trivialise this
line bundle.  We are now in the situation at the beginning of this section.

\section{The transgression for $p=2$ Faddeev's anomaly and gerbes}
\noindent
When $p=2$ we are interested in the transgression
$$
H^{3}(\A/\G)\to H^{2}_{L\G}(\G).
$$
We can regard the elements of $H^{3}(\A/\G,\Z)$ as characteristic
classes of $PU(H)$ bundles for $H$ an infinite dimensional separable
Hilbert space.  They measure the failure of the bundle to
extend to a $U(H)$ bundle and hence to be trivial as $U(H)$
is contractible.

If $Q$ is a $PU(H)$ bundle on $\A$ then it extends to a
$U(H)$ bundle because $\A$ is contractible.  However if $\G$ acts
on $Q$ then it may not act on $PU(H)$ but a central extension
will.  This central extension is determined by a class in $H^{2}_\G
(\Map(\A,U(1)))$ and determines a class on $H^{2}_{L\G}
(\Map(\A,\R))$.
We claim that these two classes are related by transgression.
To obtain a de Rham version of the three class we need to use the theory
of gerbes\cite{Brylinski} In fact we shall use a simplified definition
of a gerbe
that will be adequate for our purposes. The relationship of this
definition to that of Brylinski is similar to the relationship between
sheaves and line bundles.

\subsection{Gerbes}
\noindent
We want to define  a differential geometric object that, for want of a better
name,  we
shall call a gerbe. The reader should beware that these are not the
same as the gerbes defined by Brylinski.   We shall indicate
below the precise relationship.
Let $\pi \colon Y \to M$ be  a
fibration. Define the fibre product $Y \times_M Y$ as
the subset of pairs $(y, y')$ in
$Y \times Y$ such that $\pi(y) = \pi(y')$. Notice that the
diagonal is inside $Y\times_M Y$ and that the map
$\tau$ that transposes two elements fixes $Y \times_M Y$.  Then
a gerbe is a principal $\C^\times$ bundle $Q$ over $Y \times_M Y$
with a composition map defined as follows. Inside the product
$Y\times_M Y \times Y\times_MY $ we can consider all the pairs of
the form $((p,q), (q, r))$. We shall denote this subset by
$Y\times_M Y \circ Y\times_M Y$.  The  bundle $Q$ tensored with
itself defines a line bundle over $Y\times_M Y \times Y\times_MY $
which we shall denote by $Q\otimes Q$. Its restriction to
$Y\times_M Y \circ Y\times_M Y$ we shall denote by $L\circ L$.
The composition is a morphism of bundles $L\circ L$
which we denote by $v\otimes w \mapsto v\circ w$.
We require that the composition is associative, that
is $(v\circ w)\circ u = v\circ(w\circ u)$
whenever the triple product is defined. From the composition
we can define  an identity and inverse as follows.

The identity is a section denoted $1$ of $Q$ restricted to the
diagonal $\triangle_Y$ defined by $v\circ 1 = v$.  Notice that
this is well-defined as if $v \in Q_{pq}$ and $e \in Q_{qq}$
then $v\circ e \in Q_{pq}$ and hence $v\circ e = \alpha v$
for some $\alpha$ in $U(1)$. Hence we define $1 = e/\alpha$.
Of course we would also like that $1\circ v = v$. It
is clear that $1\circ v = \beta v$ and hence using associativity
on the product $v\otimes v$ we deduce that $\beta = 1$.

The inverse is a  bundle morphism $Q \to Q^*$ covering the
transposition map $(p,q) \to (q,p)$ on $Y\times_M Y$.
If we denote it by $v \mapsto v^{-1}$ it is defined by
$v\circ v^{-1} = 1$.

A simple example is to  let $Q \to Y$ be a principal $\C^\times$ bundle.
Define $P_{(x, y)} = Aut_{\C^\times}(Q_y, Q_x)$. Then this defines a gerb
called the trivial gerb.
More generally, associated to every gerbe is a
cohomology class in $H^3(M, \Z)$. This class
vanishes if the gerbe is trivial.

The example we are interested in as follows.  Let
$$
0 \to \C^\times \to \hat G \to G \to 0
$$
be a central extension of groups. Let $Y \to M$ be a principal
$G$ bundle. Let $p \colon \hat G \to G$ be the projection and
define a gerb over $Y \times_M Y$ by
$$
\hat Y_{(x, y)} = \{ g \in \hat G \mid x  = y p(g)\}.
$$
Assume that $Y$ has a lift to a principal $\hat G$
bundle $Z$ over $M$ so that there is a projection
$q \colon Z \to Y$ commuting with $p$ in the appropriate way.
Then $Z \to Y$ is a $\C^\times$ bundle over $Y$. Let $g \in  \hat Y_{(x, y)}$
thenit defines a map $Z_y \to Z_x$ which, by centrality, commutes
with the $\C^\times$ action. This defines an isomorphism
$$
\hat Y_{(x, y)} \simeq Aut_{\C^\times}(Z_x, Z_y)
$$
so that the gerbe $\hat Y$ is trivial. Moreover if the gerbe
$\hat Y$ is trivial, say isomorphic to $Aut(Z, Z)$ for
some $\C^\times$ bundle $Z \to Y$ then we can define an action
of $\hat G$ on $Z$ and make it a lift of $Y$.  To do this start
with $g$ in $\hat G$ and the fibre $Z_y$. Then let $x = y p(g)$.
Then $g \in \hat Y_{(x, y)} = Aut_{\C^\times}(\hat Z_y, \hat Z_x)$
so apply the corresponding automorphism to any element in $\hat Z_y$
to define the action of $g$.  It can be checked that this
defines a lift of $Y$.
In this case the gerbe class is the obstruction to the bundle lifting. The
usefulness of gerbes for our purposes is that the give a differential
form realisation of this class that can be used to investigate transgression.

To define this differential form we have to introduce the notion of a
gerbe connection. This is a connection on
$Q \to Y\times_M Y$ which respects the
structure of the gerbe. This means that over the diagonal it is the flat
connection, that the product map on the gerbe maps the product connection
to itself and that the inversion map maps the connection to its dual. The
curvature
of the gerbe connection is a two form $F$ on $Y\times_M Y$. It can be shown
that the curvature can be written as $F = \pi_1 ^* f - \pi_2^* f$
for some two-form $f$ on $Y$ where $\pi_i$ denotes the projection onto
the various factors.\cite{Murray}
 It can then be shown that $df$ is the pull-back of a
three-form $\omega$ on $M$ which we shall call the gerbe curvature.

The relationship between our line bundles and the gerbes
defined by Brylinski\cite{Brylinski} is as follows. For any open set $U
\subset M$
we can consider the space of all sections $s$ of $Y$. To
any two such sections $s$ and $t$ we have section over $U$ of
$Y\times_MY$ defined by $m \mapsto (s(m), t(m))$. We can use this
to pullback the line bundle $L$ to a line bundle on $U$.
The space of all sections of this we call $Mor(s,t)$.
There is an obvious composition of morphisms in $Mor(s,t)$
with morphisms in  $Mor(t,r)$ to give elements of $Mor(s,r)$.
We therefore have associated to $U$ a category, in fact a
groupoid.  This association is a presheaf of groupoids. If
we sheafify it we obtain sheaf of groupoids which is a gerbe.
More details can be found in the work of Murray.\cite{Murray}

\subsection{Gerbes and co-cycles.}
\noindent
Let $P \to \A$ be a fibration and $J \to P$ a gerbe. Assume that $\G$
acts on $P$ and $J$  and preserves the gerbe structure.
Choose a section $s \colon \A \to P$. This is possible because $\A$ is
contractible.  Let $\pi$ denote the projection $P \to \A$ and
define a map $\phi \colon P \to P*P$ by
$$
\phi(p) = (p, s(\pi(p))).
$$
Define a $U(1)$ bundle $Q$ on $P$ by $Q = \phi^{-1}(P)$. That
is $Q_p = J_{(p, s(\pi(p)))}$. Notice then that
$$
Q_p\otimes Q_q^* = P_{p, s(\pi(p))}\otimes P_{s(\pi(p))}^* = P_{p,q}
$$
so that $Q$ trivialises the gerbe $P$.  Consider now the automorphisms of
$Q$ covering the action of $\G$ on $P$  and commuting with
the identification of $J$ and $Q\otimes Q^*$. That is if $\hat g$ is such an
automorphism covering $g$ then we have that the diagram
$$
\matrix{
 Q_p\otimes Q_q^* & \to & J_{p,q} \cr
\hat g\otimes \hat g \downarrow \phantom{\hat g\otimes \hat g}
        & \ & g \downarrow \phantom{g} \cr
 Q_{gp}\otimes Q_{gq}^* & \to & J_{gp,gq} \cr
}
\eqno(4.1)$$
commutes.

 To see that there are any let
 $g$  be in $\G$ and consider the bundle over $\A$ whose fibre at some $A$
is $J_{gs(A), s(A)}$. Because $\A$
is contractible we can find a section $\eta$
of this. Then we can define an automorphism of  $Q$ covering $g$ by
noticing that
$g$ maps $Q_p$ to $J_{gp, gs\pi(p)}$ and that $J_{gp, gs\pi(p)} \otimes
J_{gs(\pi(p)), gs\pi(p)} = Q_{gp}$.  It is straightforward to check that
this automorphism preserves the identification $P = Q\otimes Q^*$.
Hence  if we apply $g$ and then
pair with $\eta$ we finish up in $Q_{gp}$.  Notice that any two
lifts of $g$ defined in this way differ by an element of $\Map(\A, \R)$.
Conversely any two automorphisms of $Q$ covering the action of some element
of $\G$ must differ by an automorphism fixing $P$. This is  a map
from $P$ into $U(1)$. Because of the commutativity of the diagram (4.1)
we see that this defines a map from $Q_p\otimes Q^*$ to itself which must be
the identity. Hence the map into $U(1)$ is
constant on the fibres of $P \to \A$
and defines an element of $\Map(\A, \R)$. Let $\hat \G$ be this group
of automorphisms. We have proved that there is a central extension
$$
U(1) \to \hat \G \to G.
$$

We are interested in the co-cycle for the corresponding extension
of Lie algebras.  The Lie algebra of an automorphism is a vector
field. To define explicitly a co-cycle we start we need a method
of lifting vector fields from $P$ to $Q$. If this is denoted $X \mapsto
\tilde X$
then the cocycle is
$$
c(X, Y) = [\tilde X, \tilde Y] - \widetilde([X, Y]).
$$
A method of defining such a lifting is to pick a connection $\nabla$ for $Q$.
This needs to be a connection with the property that the induced
connection on $J$ is $\G$ invariant.
We shall see that this is possible in a moment.
Given the connection we can define the lift of a vector field by lifting
it horizontally. Standard results then tell us that the cocycle is
$$
c(X, Y) = F(X, Y)
$$
where $F$ is the curvature.

Consider now the quotient gerbe over $\A/\G$. Pick a gerbe connection $\nabla$,
two-form $f$ and gerbe curvature $\omega$. Denote the corresponding
pulled-back objects by $\hat \nabla$, $\hat f$ and $\hat \omega$. Notice,
that by definition they are $\G$ invariant. We can then construct a connection
on $Q$ of the required type as $\phi^*(\hat \nabla)$. The curvature of this
is $\phi*(\pi_1^* + \pi_2^*)(\hat f)$. Hence the cocycle is
$$
\begin{array}{ll}
c(X, Y) &= \hat f_{s\pi(p)}(X, Y) - \hat f_p(s_*\pi_*(X),s_*\pi_*(Y))\cr
    &= \hat f_{s\pi(p)}(X, Y)\cr
 \end{array}
$$
as $\hat f$ is pulled-back under $\G$ and $X$ and $Y$ are tangent to
an orbit of $\G$.   But we also have $p^*(\omega) = ds^*\hat f$
so that the class defined by the transgression is
$$
s^*\hat f(X, Y) = \hat f_{s\pi(p)}(X, Y)
$$
which is the same as the cocycle for the extension.

\subsection{The Fock bundle and the Faddeev-Mickelsson anomaly}
\noindent
Let $M$ be odd dimensional so that the Dirac operator
maps from spinors to spinors.   Following\cite{Segal}
consider the subset $\A_0 \subset \A \times \R$ defined as all
pairs $(A, s)$ where $s$ is not in the spectrum of the Dirac operator
coupled to $A$.  Given such a pair $(A, s)$ we can decompose the
Hilbert space of spinors into the direct sum of $H^+_{(A, s)}$
and $H^-_{(A, s)}$; the sums of the eigenspaces for eigenvalues
greater and less than $s$ respectively. We can then form the Fock
bundle
$$
F_{(A, s)} = \bigwedge(H^+_{(A, s)}) \otimes (\bigwedge(H^-_{(A, s)}))^*.
$$
If leave $A$ fixed and vary $s$ to some $t< s$. Then we have
$$
H = (H^-_{(A, t)}) \oplus V_{(A,t,s)} \oplus  H^+_{(A, s)}
$$
where $V_{(A,t,s)}$ is the sum of all the eigenspaces for eigenvalues
between $t$ and $s$. Moreover
$$
H^+_{(A, t)} =  V_{(A,t,s)} \oplus H^+_{(A, s)} \quad{\rm and}\quad
H^-_{(A, s)}  = H^-_{(A, t)} \oplus V_{(A,t,s)}.
$$
It follows that
$$
F_{(A,s)} = F_{(A,t)}\otimes L_{(A, s,t)}
$$
where $L_{(A,s,t)} = (\det V_{(A,t,s)})^2$.  Hence the projective spaces
$P(F_{(A,t)})$ and $P(F_{(A,t)})$ can be identified and descend to a
projective bundle on $\A$.  Moreover the gauge group acts on the projective
bundle and we are therefore in the setting of section 4.

\section{3-cocycles}
\noindent
The main examples of 3-cocycles which have occurred in the physics
literature do not fit into
transgression map framework that we have presented and  they are not
given by cocycles on the Lie algebra of the gauge group.
Moreover they
have been interpreted as the breakdown of the Jacobi identity. However
algebras of operators on Hilbert spaces must
always satisfy the Jacobi identity
and so this forces the conclusion that when a 3-cocycle arises
it must signal the absence of a representation by operators of the
algebra or group in question.
Some time ago one of us\cite{Carey} showed how one could understand this
conclusion in terms of an `obstruction' to representing a symmetry
group of a quantum system on a Hilbert space (this was an extension of the
traditional
mathematical approach to the existence of group extensions\cite{Maclane}).
Recently\cite{CGRS} a mathematical
framework,
 originally devised to study group actions on C$^*$-algebras,
was adapted to understand the examples of
3-cocycles arising in quantum field theory.
This approach was applied in\cite{CGHL} to study 3-cocycles in non-abelian
gauge theories.
It seems plausible that the geometric framework of this paper
can be applied to these examples as well although we have not yet found a
way to do so.

\newpage
\subsection{Summary of Jo's Calculation}
\noindent
We begin by reminding the reader of the standard calculation in
the physics literature following Jo.\cite{Jo}
Jo considers chiral fermions in $(3+1)$ dimensions coupled to a Yang-Mills
gauge field.
He constructs an equal-time algebra of quantum fields starting with
$$
A=\sum_{i=1}^{3}\;\sum_{a}\;A_{i}^{a}(x)T^{a}\,dx^{i},
$$
the Yang-Mills field ``operator'' at fixed time $t=0$. Here $T^{a}$ are the
generators of
the Lie algebra ${\bf g}$ of the gauge group. Jo finds that the CCR become
anomalous. Specificaly he calculates commutation relations for
$A$ and its conjugate momentum $E$ by a perturbation theory method
(the BJL method\cite{jackiw}.) They are:
$$
\begin{array}{l}
[A_{i}^{a}({x}),A_{j}^{b}({y})]=\emptyset \\[+10pt]
[E_{i}^{a}({x}),A^{b}_{j}({y})]=-i\delta_{ij}\delta^{ab}\delta^{3}
({x}-{y}) \\[+10pt]
[E_{i}^{a}({x}),E_{j}^{b}({y})]=i\alpha\epsilon_{ijk}\mbox{ tr}[(T^{a}T^{b}+
T^bT^a)A_{k}({x})]\delta^{3}({x}-{y})
\end{array}
$$
(here $\epsilon_{ijk}$ is the
antisymmetric tensor as usual, tr$(\cdot)$ is the trace in ${\bf g}$, $\alpha$
is a constant,
$\displaystyle A_{k}\equiv\sum_{a}A_{k}^{a}T^{a}$ and ${\bf g}$ is
$n$-dimensional).

As it stands Jo's calculation is not in the right form for our purposes.
Let ${\cal S}$ denote the smooth functions of fast decrease on $\R^{3}$
with values in $\R^{3}\times \R^{n}$, and denote the components of
$f\in{\cal S}$ by $f^{a}_{i}(x)$.  Then
$$
A(f):=\sum_{i=1}^{3}\;\sum_{a=1}^{n}\int_{\R^{3}}A^{a}_{i}(x)f^{a}_{i}(x)d^{3}
x\eqno(5.1)
$$
and similarly we smear $E(f)$ over ${\cal S}$.  The commutation relations are:
$$
[A(f),A(g)]=0\eqno(5.2\mbox{i})
$$
$$
[E(f),A(g)]=-i\sum_{j=1}^{3}\;\sum_{a=1}^{n}\int f^{a}_{j}(x)g^{a}_{j}(x)d^
{3}(x)=:-i(f,g)\eqno(5.2\mbox{ii})
$$
$$
[E(f),E(g)]=i\alpha A(f  \star \;\; g)\eqno(5.2\mbox{iii})
$$
where
$$
\displaystyle(f \star
\;\;g)^{c}_{k}(x):=\sum_{a,b}^{n}\epsilon^{ijk}f^{a}_{i}(x)
g^{b}_{J}(x)\;tr[(T^{a}T^{b}+T^{b}T^{a})T^{c}].
\eqno(5.3)
$$

The right-hand side of (5.2iii) is not a
2-cocycle, only a 2-cochain with values
in the Lie algebra generated by the $A(f)$.
Now proceeding as in Jo\cite{Jo} with these smeared commutation relations, we
find that the triple commutators produce a scalar-valued three-cocycle:
$$
\begin{array}{llr}
J(E(f),E(g),E(h))
&=\mbox{Cycl.}[E(f),[E(g),E(h)]] &\\
&=\mbox{Cycl.}[E(f),i\alpha A(g \star \;\;h)] &\\
&=\alpha\mbox{ Cycl.}(f,g \star \;\;h) &\\
\displaystyle&=\alpha\mbox{ Cycl.}\sum_{a,b,c}\int\epsilon^{ijk}g^{a}_{i}(x)
h^{b}_{j}(x)f^{c}_{k}(x)\;tr[(T^{a}+ \dots &\\
&        \quad\quad\quad \dots T^{b}+T^{b}+T^{a})T^{c}]d^{3}x &\\
\displaystyle&=3\alpha\sum_{a,b,c}\int\epsilon^{ijk}g^{a}_{i}(x)h^{b}_{j}(x)f^
{c}_{k}(x)\;tr[(T^{a}T^{b}+ \dots &\\
& \quad\quad T^{b}T^{a})T^{c}]d^{3}x &(5.4)\\
\end{array}
$$
where Cycl. denotes summation over cyclic permutations of $f,g,h$.

At this point the view expounded in the literature is to regard the Jacobi
identity as failing.
So one has to conclude that the $E(f)$'s
are not operators on a common dense invariant domain. In
fact of course it is not clear where the contradiction lies since one
assumes the $E(f)$'s are such  operators
in order to define the anomalous commutators (5.2) and so perhaps the
BJL method is the real problem. There is a way of making sense of all this
which is expounded in\cite{CGHL} following the framework
of\cite{CGRS} and which we will now summarise.

\subsection{3-cocycles as obstructions}
\noindent
To explain this we need to identify an exact sequence of Lie algebras
$$
0\rightarrow \Delta\rightarrow\Gamma\rightarrow\Omega\rightarrow 0
\eqno(5.5)
$$
with $\Delta$ an Abelian Lie algebra and $\Gamma$ an Abelian extension of
$\Delta$ by $\Omega$.

To obtain this sequence, we will think of the ``operators'' $E(f),A(g)$
as acting by derivations on the algebra ${\cal A}$
generated by the space time smeared fields, in which case scalars are
factored out of the commutation relations (5.2). (One would
calculate this action by taking formal commutators of elements of ${\cal
A}$ with the
$E(f), A(g)$. This is not inconsistent because even though
$E(f)$ may not exist as an operator one only needs
it to specify a quadratic form in order that
the formal commutator with elements of $\cal A$ should be defined.)
Let $\Delta$ be the Abelian Lie algebra, identical with the test function
space ${\cal S}$ as a linear space (its elements are thought of as the
$A(f)$'s).  Let $\Omega$ be another copy of ${\cal S}$ as an Abelian Lie
algebra (but its elements are thought of as the $E(f)$'s acting by
commutators so the
right-hand side of (3.2iii) does not enter).

The action of $\Omega$ on $\Delta$ is taken to be trivial,  (this is
justified by
(5.2ii), factoring out the constants).  Then the map $\sigma:\Omega^{2}
\rightarrow\Delta$ defined by $\sigma(f,g):=\alpha f \star
\;\;g$ (using the identification of $\Omega$ and $\Delta$ with ${\cal S}$)
is a two-cycle because
$$
(\partial\sigma)(f,g,h)=\mbox{Cycl.}\{d_{f}(\sigma(g,h))+\sigma(f,[g,h]))\}=0
$$
for all $f,g,h\in\Omega$.  So we form the corresponding extension of Lie
algebras:
$$
\Gamma=\Omega\oplus\Delta\;\;\mbox{ with bracket }\;\;[f\oplus g,\;h\oplus k]
=0\oplus\sigma(f,h)
\eqno(5.6)
$$
  Now that the  exact sequence (5.3) is specified, and we
identify a section $\omega:\Omega\rightarrow\Gamma$ by $w_{f}=f\oplus 0$,
so
$
[\omega_{f},\omega_{h}]=0\oplus\sigma(f,h),
$
and $\omega_{f}$ corresponds to $E(f)$, now satisfying (5.2iii).

We now aim to construct a  second exact sequence
$$
0\to{\cal W}\to{\cal V}\;\stackrel{ad|_{\cal A}}
{\to}\;
\Delta\to 0
$$
by assuming there is a representation
$v$:$\Delta\rightarrow {\cal V}$ by self-adjoint
operators on
a common dense invariant domain $D$ in a Hilbert space ${\cal H}$, on which
the algebra $\cal A$ is also irreducibly represented and such that
$v$ implements $\Delta$ as derivations on
${\cal A}$ (that is the action of $d\in\Delta$
on $\cal A$ is given by
$ad\vert_{\cal A}(d)(A)=[v_d, A]$ for $A\in \cal A$). Let ${\cal V}$ be
the abelian Lie algebra given by the linear span
of the operators $\{I\!\!R 1\!\!1, v_{d}\;|\;d\in\Delta\}$.
Then ${\cal W}=I\!\!R 1\!\!1$ and $ad|_{\cal A}({\cal V})\cong\Delta$.

The framework in\cite{CGRS} requires us to construct an intermediate
invariant whose geometric significance is completely opaque.
This invariant is constructed from
a map $\lambda:\Gamma\times\Delta\rightarrow I\!\!R$
determined by an  action $\delta:\Gamma\rightarrow\mbox{Der }
{\cal V}$. The relation (5.2ii) which is suggested by the canonical
commutation relations
is what we use to define $\lambda$.
Replace $A(f)$ with $v_{f}$, $f\in\Delta={\cal S}$
and suppose for the moment that there are
operators
$u_{f}$, $f\in\Omega={\cal S}$ preserving $D$
(these would be the  $E(f)$'s if they existed)
satisfying the usual commutation relations
$
[u_{f},v_{h}]=i(f,h)
$.
Suppose further that these commutation relations
provide the action $\delta:\Omega\rightarrow\mbox{Der }{\cal V}$ by
$d_{f}(v_{h})=(-i)[u_{f},v_{h}]$.  So $\delta:\Gamma\rightarrow\mbox{Der }
{\cal V}$ will be
$$
\delta_{f\oplus g}(v_{h})=\delta_{f\oplus 0}(v_{h})+\delta_{0\oplus g}(v_{h})
=\delta_{f\oplus 0}(v_{h})=(f,h)
$$
This tells us to define  $\lambda:\Gamma\times\Delta\rightarrow I\!\!R$  by
using\cite{CGHL}
$$
\begin{array}{rlr}
\lambda(f\oplus g,h) &=
\delta_{f\oplus g}(v_{h})-g_{[f\oplus g,0\oplus h]} &\\
&=\delta_{f\oplus g}(v_{h}) &\\
 &= (f,h).&\hfill(5.7)
\end{array}
$$
 Now this method of arriving at (5.7) is not an argument because the
$u_f$'s need not exist however (5.7) is a good definition provided\cite{CGRS}
$$
\lambda([f\oplus g,h\oplus k],m)=\lambda(f\oplus g,[h\oplus k,0\oplus m])-
\lambda(h\oplus k,[f\oplus g,0\oplus m])
$$
which is easy to check.

Hence the $\lambda$ given by (5.7) satisfies the
requirements of\cite{CGRS} and
hence defines a 3-cocycle
$K:\Omega^{3}\rightarrow \R:$
$$
\begin{array}{rlr}
K(f,g,h)  &=  \lambda(f\oplus 0,\sigma(g,h))+\lambda(g\oplus 0,\sigma(h,f))+
\lambda(h\oplus 0,\sigma(f,g)) &\\
 &=  (f,\sigma(g,h))+(g,\sigma(h,f))+(h,\sigma(f,g)) &\\
&=\alpha\mbox{ Cycl.}(f,g \star \;\;h)    &(5.8)\\
\end{array}
$$
which is exactly the same 3-cocycle as the one obtained in
Jo's calculation.  In\cite{CGHL}
it is shown that the Lie algebra $\Gamma$ cannot be represented in such a way
that the
action $\delta$ associated to $\lambda$ can be implemented
because
$K$ determines a non-trivial cohomology class.
(The cohomology group with complex coefficients in
degree $n$ of
an abelian Lie algebra is given by the space of totally skew
$n$-multilinear maps).

{}From the viewpoint of quantum field theory this is
further evidence that the canonical equal-time formalism is probably not
appropriate in 3+1 dimensions: a fact which is not all that surprising.

The relationship with our earlier discussion arises by
letting $\A$ be the connections on a trivial bundle over $\R^3$ (i.e at
fixed time
$t=0$) and Map$(\A,\C)$ the Abelian Lie algebra of
continuously differentiable mappings $\A\to \C$ (so that
$\delta d/\delta A(f)$ makes sense $\forall\;d\in\Map(\A, \C)$).
Then we identify
$\Delta$ as a subgroup of $ \Map(\A,\C)/\C$
by regarding each $f\in \Delta$  as defining a function $A(f)$ on $\A$
via (5.1).
However
it is not clear what the geometric meaning of this $3$-cocycle is.
There are generalisations of categories to a theory of $2$-categories,
where one has objects,  morphisms between pairs of objects and something
else between any triangle of morphisms. Perhaps  the correct geometric
object is a sheaf of $2$-categories?

\nonumsection{Acknowledgements}

For the results described in the last section we thank our collaborators
for their assistance. We are indebted in a more fundamental way
to Bert Green and Angas Hurst for helping
to create the intellectual environment
in which these investigations could be undertaken.

\nonumsection{References}

\end{document}